\documentstyle[12pt]{article}

\font\twlgot =eufm10 scaled \magstep1 \font\egtgot =eufm8
\font\sevgot =eufm7 \font\twlmsb =msbm10 scaled \magstep1
\font\egtmsb =msbm8 \font\sevmsb =msbm7

\newfam\gotfam
\def\pgot{\fam\gotfam\twlgot}
\textfont\gotfam\twlgot \scriptfont\gotfam\egtgot
\scriptscriptfont\gotfam\sevgot
\def\got{\protect\pgot}
\newfam\msbfam
\textfont\msbfam\twlmsb \scriptfont\msbfam\egtmsb
\scriptscriptfont\msbfam\sevmsb
\def\Bbb{\protect\pBbb}
\def\pBbb{\relax\ifmmode\expandafter\Bb\else\typeout{You cann't use
Bbb in text mode}\fi}
\def\Bb #1{{\fam\msbfam\relax#1}}

\newcommand{\gd}{{\got d}}

\newcommand{\ccG}{{\got g}}

\def\op#1{\mathop{\fam0 #1}\limits}

\newcommand{\id}{{\rm Id\,}}

\newcommand{\nm}[1]{|{#1}|}
\newcommand{\beq}{\begin{equation}}
\newcommand{\eeq}{\end{equation}}
\newcommand{\ben}{\begin{eqnarray}}
\newcommand{\een}{\end{eqnarray}}
\newcommand{\be}{\begin{eqnarray*}}
\newcommand{\ee}{\end{eqnarray*}}
\newcommand{\bea}{\begin{eqalph}}
\newcommand{\eea}{\end{eqalph}}

\newcommand{\cR}{{\cal R}}

\newcommand{\cH}{{\cal H}}

\newcommand{\cS}{{\cal S}}

\newcommand{\sG}{{\cal G}}

\newcommand{\al}{\alpha}

\newcommand{\bt}{\beta}
\newcommand{\dl}{\delta}
\newcommand{\la}{\lambda}
\newcommand{\La}{\Lambda}
\newcommand{\f}{\phi}
\newcommand{\om}{\omega}

\newcommand{\m}{\mu}

\newcommand{\g}{\gamma}

\newcommand{\vf}{\varphi}

\newcommand{\di}{{\rm dim\,}}
\newcommand{\pr}{{\rm pr}}

\newcommand{\si}{\sigma}
\newcommand{\Si}{\Sigma}
\newcommand{\w}{\wedge}

\newcommand{\wh}{\widehat}
\newcommand{\ol}{\overline}
\newcommand{\dr}{\partial}
\newcommand{\ar}{\op\longrightarrow}

\newcommand{\ot}{\otimes}

\newcommand{\ve}{\varepsilon}

\newcounter{eqalph}
\newcounter{equationa}
\newcounter{remark}
\newcounter{example}
\newcounter{theorem}
\newcounter{proposition}
\newcounter{lemma}
\newcounter{corollary}
\newcounter{definition}
\def\theremark{\arabic{remark}}
\def\thetheorem{\arabic{theorem}}

\def\thedefinition{\arabic{definition}}

\newenvironment{theo}{\refstepcounter{theorem}
\bigskip\noindent{\bf Theorem \thetheorem.} \it}{\medskip}

\newenvironment{defi}{\refstepcounter{definition}
\bigskip\noindent{\bf Definition \thedefinition.}\it}{\medskip}

\newenvironment{eqalph}{\stepcounter{equation}
\setcounter{equationa}{\value{equation}} \setcounter{equation}{0}

\begin{eqnarray}}{\end{eqnarray}
\setcounter{equation}{\value{equationa}}}

\newcommand{\mar}[1]{}

\begin{document}
\hbox{}

{\parindent=0pt

{\large\bf Reduction of principal superbundles, Higgs superfields,
and supermetric}
\bigskip

{\bf G Sardanashvily}

\medskip
{\small

Department of Theoretical Physics, Moscow State University, 117234
Moscow, Russia

%\medskip

%E-mail: sard@grav.phys.msu.su

\bigskip

{\bf Abstract}

By virtue of the well-known theorem, a structure Lie group $G$ of
a principal bundle $P\to X$ is reducible to its closed subgroup
$H$ iff there exists a global section of the quotient bundle
$P/H\to X$. In gauge theory, such sections are treated as
classical Higgs fields, and are exemplified by Riemannian and
pseudo-Riemannian metrics. This theorem is extended to a certain
class of principal superbundles, including a graded frame
superbundle with a structure general linear supergroup. Each
reduction of this structure supergroup to an orthgonal-symplectic
supersubgroup is associated to a supermetric on a base
supermanifold.

\bigskip

% PACS numbers:

} }

\section{Introduction}

In gauge theory on a principal bundle $P\to X$ with a structure
Lie group $G$ reduced to its subgroup $H$, the corresponding
global section of the quotient bundle $P/H\to X$ (see Theorem
\ref{g00} below) is regarded as a classical Higgs field
\cite{nik,keyl,sard92,higgs}, e.g., a gravitational field in gauge
gravitation theory \cite{iva,tmf,pref}. A particular local case of
this construction is the so called nonlinear realization of a
group $G$ extending a linear representation of its Cartan subgroup
$H$ \cite{col,jos}. In fact, it is a representation of the Lie
algebra of $G$ around its origin (see Appendix). Nonlinear
realization of different groups have been studied, e.g., in
application to gravitation theory \cite{ish,tiembl,kirsch,lecl}.

SUSY gauge theory including supergravity is greatly motivated by
grand-unification models and contemporary string and brane
theories. It is mainly developed as a Yang--Mills type theory with
spontaneous breaking of supersymmetries
\cite{niew,nill,west,wess,binet}. There are various
superextensions of the pseudo-orthogonal and Poincar\'e Lie
algebras \cite{aleks,aur,aur2,aleks2}. The nonlinear realization
of some Lie superalgebras have been studied
\cite{volk,ivanov,shima,clark}. However, supergravity introduced
in SUSY gauge theory has no geometric feature, e.g., as a
supermetric \cite{iva86,del}. Therefore, our goal here is to
extend the above mentioned Theorem \ref{g00} on the reduction of
principal bundles to principal superbundles.

Let $\pi:P\to X$ be a principal smooth bundle with a structure Lie
group $G$. Let $H$ be a closed (consequently, Lie) subgroup of
$G$. Then $G\to G/H$ is an $H$-principal fiber bundle and, by the
well known theorem, $P$ is split into the composite fiber bundle
\mar{g1}\beq
P\ar^{\pi_H} P/H\ar X, \label{g1}
\eeq
where $P\to P/H$ is an $H$-principal bundle and $P/H\to X$ is a
$P$-associated bundle with the typical fiber $G/H$ \cite{ste}. One
says that the structure group $G$ of a principal bundle $P$ is
reducible to $H$ if there exists an $H$-principal subbundle of
$P$. The necessary and sufficient conditions of the reduction of a
structure group are stated by the well known theorem \cite{ste}.

\begin{theo} \label{g00} \mar{g00} There is one-to-one
correspondence $P^h=\pi_H^{-1}(h(X))$ between the reduced
$H$-principal subbundles $P^h$ of $P$ and the global sections $h$
of the quotient bundle $P/H\to X$.
\end{theo}

As was mentioned above, sections of $P/H\to X$ are treated in
gauge theory as classical Higgs fields. For instance, let $P=LX$
be the $GL(n,\Bbb R)$-principal bundle of linear frames in the
tangent bundle $TX$ of $X$ (n=\di X). If $H=O(k,n-k)$, then a
global section of the quotient bundle $LX/O(k,n-k)$ is a
pseudo-Riemannian metric on $X$.

Let $V$ be a vector space, and let a group $G$ acts on $V$ on the
left. Then the quotient
\mar{g2}\beq
E=(P\times V)/H \label{g2}
\eeq
is a fiber bundle associated with the $H$-principal bundle $P\to
P/H$. The quotient (\ref{g2}) is defined by identification of the
elements $(p,v)$ and $(ph,h^{-1}v)$ for all $h\in H$. It is a
composite bundle
\mar{g3}\beq
E\to P/H\to X, \label{g3}
\eeq
with the typical fiber $G/H\times V$. Sections of $E\to X$
(\ref{g3}) are the pairs of matter and Higgs fields. It should be
emphasized that $E\to X$ fails to be a fiber bundle with the
structure group $G$, unless the $H$-principal bundle $G\to G/H$ is
trivial. If $G\to G/H$ is a trivial bundle, there exists its
global section whose values are representatives of elements of
$G/H$. Given such a section $s$, the typical fiber of $G/H\times
V$ of $E\to X$ can be provided with the particular induced
representation
\mar{g4}\beq
G\ni g:(\si,v) \mapsto (g\si, g_\si v), \qquad
g_\si=s(g\si)^{-1}gs(\si)\in H, \label{g4}
\eeq
of $G$. Of course, this representation is not canonical, unless
$V$ itself admits a representation of $G$. If $H$ is a Cartan
subgroup of $G$, the above mentioned nonlinear realization of $G$
in a neighbourhood of its unit exemplifies the induced
representation (\ref{g4}) (see Appendix).

Our goal here is the following extension of Theorem \ref{g00} to
principal superbundles. We consider Lie supergroups and principal
superbundles in the category of $G$-supermanifolds \cite{bart}.

\begin{theo} \label{g20} \mar{g20}
Let $\wh P\to \wh M$ be a principal $G$-superbundle with a
structure $G$-Lie supergroup $\wh G$, and let $\wh H$ be a closed
$G$-Lie supersubgroup of $\wh G$ such that $\wh G\to \wh G/\wh H$
is a principal superbundle. There is one-to-one correspondence
between the principal $G$-supersubbundles of $\wh P$ with the
structure $G$-Lie supergroup $\wh H$ and the global sections of
the quotient superbundle $\wh P/\wh H\to \wh M$ with the typical
fiber $\wh G/\wh H$.
\end{theo}

In order to proof Theorem \ref{g20}, it suffices to show that the
morphisms
\mar{g21}\beq
\wh P\ar \wh P/\wh H\ar \wh M \label{g21}
\eeq
form a composite $G$-superbundle. A key point is that underlying
spaces of $G$-supermanifolds are smooth real manifolds, but
possessing very particular transition functions and morphisms.
Therefore, the condition of local triviality of the quotient $\wh
G\to \wh G/\wh H$ is rather strong. However, it is satisfied in
the most interesting case for applications when $\wh G$ is a
supermatrix group and $\wh H$ is its Cartan supersubgroup. For
instance, let $\wh P=L\wh M$ be a principal superbundle of graded
frames in the tangent superspaces over a supermanifold $\wh M$ of
even-odd dimension $(n,2m)$. If its structure general linear
supergroup $\wh G=\wh{GL}(n|2m; \La)$ is reduced to the
orthgonal-symplectic supersubgroup $\wh H=\wh{OS}p(n|m;\La)$, one
can think of the corresponding global section of the quotient
bundle $L\wh M/\wh H\to \wh M$ as being a supermetric on $\wh M$.
Note that a Riemannian supermetric on graded manifolds has been
considered in a different way \cite{zirn,goert}.

It should be emphasized that there are different notions of a
supermanifold, Lie supergroup and superbundle. Let us mention a
definition of a super Lie group as a Harish--Chandra pair of a Lie
group and a super Lie algebra \cite{del,carm}. We are not
concerned with graded manifolds \cite{ber,kost77,man}, graded Lie
groups \cite{alm,boy}, and graded principal bundles
\cite{alm,stavr}. It should be emphasized that graded manifolds
are not supermanifolds in a strict sense. For instance, the
tangent bundle of a graded manifold fails to be a graded manifold.
However, every graded manifold can be associated to a DeWitt
$H^\infty$-supermanifold, and {\it vice versa} \cite{bart,batch2}.

We here restrict our consideration to supermanifolds over
Grassmann algebras $\La$ of finite rank. This is the case of
smooth ($GH^\infty$-, $H^\infty$-, $G^\infty$-) supermanifolds and
$G$-supermanifolds \cite{bart}. By analogy with manifolds, smooth
supermanifolds are constructed by gluing of open subsets of
supervector spaces $B^{n,m}$ endowed with the Euclidean topology.
If a supervector space $B^{n,m}$ is provided with the
non-Hausdorff DeWitt topology \cite{dew}, we are in the case of
DeWitt supermanifolds.

In a more general setting, one considers supermanifolds over the
so called Arens--Michael algebras of Grassmann origin, most
suitable for superanalysis and supergeometry. They are $R$- and
$R^\infty$-supermanifolds obeying a certain set of axioms
\cite{roth,bart93,bruz99}. In the case of finite Grassmann
algebra, the category of $R^\infty$-supermanifolds is equivalent
to the category of $G$-supermanifolds. In comparison with smooth
supermanifolds, $G$-supermanifolds have some important advantages
from the differential geometric viewpoint \cite{bart}. Firstly,
derivations of the structure sheaf of a $G$-supermanifold $\wh M$
constitute a locally free sheaf, which is the structure sheaf of
some $G$-superbundle $T\wh M$ regarded as a tangent superbundle of
$\wh M$. Secondly, the category of $G$-supervector bundles is
equivalent to a certain category of locally free sheaves of finite
rank just as it takes place in the case of smooth vector bundles.

\section{Supermanifolds}

This Section collects the relevant material on the graded tensor
calculus, superfunctions, and supermanifolds \cite{bart,book00}.

Let $V$ be a real vector space. Its exterior algebra
\be \La=\La_0\oplus\La_1=\w V=\Bbb R\op\oplus_{k=1} \op\w^k V
=(\Bbb R\op\oplus_{k=1} \op\w^{2k} V)\oplus(\op\oplus_{k=1}
\op\w^{2k-1} V)
\ee
is a graded commutative ring of rank $N=\di V$. We consider
Grassmann algebras of this type. Note that there is a different
definition of a Grassmann algebra, which is equivalent to the
above one only in the case of an infinite-dimensional vector space
$V$ \cite{jad}. A Grassmann algebra admits the splitting
\mar{+11}\beq
\La=\Bbb R\oplus\cR =\Bbb R\oplus\cR_0\oplus\cR_1=\Bbb R\oplus
(\La_1)^2\oplus \La_1 \label{+11}
\eeq
where $\cR$ is the ideal of nilpotents of $\La$. It is a unique
maximal ideal of $\La$, i.e., $\La$ is a local ring. The
corresponding projections $\si:\La\to\Bbb R$ and $s:\La\to\cR$ are
called the body and soul maps, respectively. Given a basis
$\{c^i\}$ for the vector space $V$, elements of the Grassmann
algebra take the form
\be
a=\op\sum_{k=0}^N \frac{1}{k!}a_{i_1\cdots i_k}c^{i_1}\cdots
c^{i_k}.
\ee
A Grassmann algebra $\La$ is a graded commutative Banach ring with
respect to the norm
\be
\|a\|_\La=\op\sum_{k=0}^N \op\sum_{(i_1\cdots i_k)}\frac{1}{k!}
\nm{a_{i_1\cdots i_k}}.
\ee
This norm provides $\La$ with the Euclidean topology of a
$2^N$-dimensional real vector space.

Let $B=B_0\oplus B_1=\Bbb R^n\oplus\Bbb R^m$ be an
$(n,m)$-dimensional graded vector space. Given a Grassmann algebra
$\La$ of rank $N$, the $\La$-envelope of $B$ is a free graded
$\La$-module
\mar{+70}\beq
B^{n|m}=\La B=(\La B)_0\oplus (\La B)_1=(\La_0^n \oplus
\La^m_1)\oplus (\La_1^n\oplus \La_0^m), \label{+70}
\eeq
of rank $n+m$. It is called the superspace. Its even part
\mar{g11}\beq
 B^{n,m}= \La_0^n \oplus \La^m_1 \label{g11}
\eeq
is a $\La_0$-module called the $(n,m)$-dimensional supervector
space. In accordance with the decomposition (\ref{+11}), any
element $q\in B^{n,m}$ is uniquely split as
\be
 q=x+y=(\si(x^i) + s(x^i))e^0_i + y^je^1_j,
\ee
where $\{e^0_i,e^1_j\}$ is a basis for $B$ and $\si(x^i)\in \Bbb
R$, $s(x^i)\in \cR_0$, $y^j\in\cR_1$. The corresponding body and
soul maps read
\be
\si^{n,m}: B^{n,m}\to \Bbb R^n, \qquad s^{n,m}: B^{n,m}\to
\cR^{n,m}=\cR^n_0\oplus \cR^m_1.
\ee
A supervector space $B^{n,m}$ is provided with the Euclidean
topology of a $2^{N-1}(n+m)$-dimensional real vector space.

The superspace $B^{n|m}$ seen as a $\La_0$-module is isomorphic to
the supervector space $B^{n+m,n+m}$. Any $\La$-module endomorphism
of $B^{n| m}$ is represented by an $(n+ m)\times (n+m)$-matrix
\mar{+200}\beq
L=\left(
\begin{array}{cc}
L_1 & L_2 \\
L_3 & L_4
\end{array}
\right) \label{+200}
\eeq
with entries taking values in $\La$. It is called a supermatrix. A
supermatrix $L$ is even (resp. odd) if $L_1$ and $L_4$ have even
(resp. odd) entries, while $L_2$ and $L_3$ have the odd (resp.
even) ones. Endowed with this gradation, supermatrices
(\ref{+200}) make up a $\La$-graded algebra. A supermatrix $L$
(\ref{+200}) is invertible iff the real matrix $\si(L)$ is
invertible. Invertible supermatrices constitute a general linear
graded group $GL(n|m;\La)$.

Turn now to the notion of a superfunction. Let $B^{n,m}$ be a
supervector space (\ref{g11}), where $\La$ is a Grassmann algebra
of rank $0<N\geq m$. Let $\La'$ be a Grassmann subalgebra of $\La$
of rank $N'$ whose basis $\{c^a\}$ is a subset of a basis for
$\La$. Furthermore, $\La$ is regarded as a $\La'$-algebra. Given
an open subset $U\subset \Bbb R^n$, let us consider a
$\La'$-valued graded function
\mar{+12}\beq
f(z)=\op\sum_{k=0}^{N'} \frac1{k!}f_{a_1\ldots
a_k}(z)c^{a_1}\cdots c^{a_k}, \label{+12}
\eeq
on $U$ with smooth real coefficients $f_{a_1\cdots a_k}\in
C^\infty(U)$. It is prolonged onto $(\si^{n,0})^{-1}(U)\subset
B^{n,0}$ as the Taylor series
\mar{+14}\beq
f(x)=  \op\sum_{k=0}^{N'} \frac1{k!}\left[
 \op\sum_{p=0}^N\frac{1}{p!}\frac{\dr^qf_{a_1\ldots a_k}}{\dr
z^{i_1}\cdots \dr z^{i_p}}(\si(x))s(x^{i_1})\cdots
s(x^{i_p})\right]c^{a_1}\cdots c^{a_k}. \label{+14}
\eeq
Then a superfunction $F(q)=F(x,y)$ on $(\si^{n,m})^{-1}(U)\subset
B^{n,m}$ is defined as the sum
\mar{+13}\beq
F(x,y)= \op\sum_{r=0}^N \frac1{r!} f_{j_1\ldots
j_r}(x)y^{j_1}\cdots y^{j_r}, \label{+13}
\eeq
where $f_{j_1\ldots j_r}(x)$ are graded functions (\ref{+14}). The
representation of a superfunction $F(x,y)$ by the sum (\ref{+13})
however need not be unique. The germs of superfunctions
(\ref{+13}) constitute the sheaf $S_{N'}$ of graded commutative
$\La'$-algebras on $B^{n,m}$. At the same time, one can think of a
superfunction (\ref{+13}) as being a smooth map of a
$2^{N-1}(n+m)$-dimensional Euclidean space $B^{n,m}$ to the
$2^N$-dimensional one $\La$.

Using the representation (\ref{+13}), one can define derivatives
of superfunctions as follows. Given a superfunction $f(x)$
(\ref{+14}) on $B^{n,0}$, its derivative with respect to an even
argument $x^i$ is defined in a natural way as
\mar{+16}\beq
\dr_if(x)= \op\sum_{k=0}^{N'} \frac1{k!}\left[
 \op\sum_{p=0}^N\frac{1}{p!}\frac{\dr^{p+1}f_{a_1\ldots a_k}}{\dr
z^i\dr z^{i_1}\cdots \dr z^{i_p}}(\si(x))s(x^{i_1})\cdots
s(x^{i_p})\right]c^{a_1}\cdots c^{a_k}. \label{+16}
\eeq
This even derivative is extended to superfunctions $F$ on
$B^{n,m}$ in spite of the fact that the representation (\ref{+13})
is not unique. However, the definition of odd derivatives of
superfunctions is more intricate.

Let $S_{N'}^0\subset S_{N'}$ be the subsheaf of superfunctions
$F(x,y)=f(x)$ (\ref{+14}) independent of the odd arguments $y^j$.
Let $\w\Bbb R^m$ be the Grassmann algebra generated by the
elements $(a^1,\ldots, a^m)$. The expression (\ref{+13}) implies
the sheaf epimorphism
\mar{+15}\be
\la: S^0_{N'}\ot\w\Bbb R^m \to S_{N'}, \qquad \op\sum_{r=0}^m
\frac1{r!} f_{j_1\ldots j_r}(x)\ot(a^{j_1}\cdots a^{j_r})\to
\op\sum_{r=0}^m \frac1{r!} f_{j_1\ldots j_r}(x)y^{j_1}\cdots
y^{j_r}. \label{+15}
\ee
This epimorphism is injective and, consequently, is an isomorphism
iff
\mar{+20}\beq
N-N'\geq m. \label{+20}
\eeq
If the condition (\ref{+20}) holds, the representation of any
superfunction $F$ by the sum (\ref{+13}) is unique, and $F$ is an
image of some section $f_\iota\ot a^\iota$ of the sheaf
$S^0_{N'}\ot\w\Bbb R^m$. Then the odd derivative of $F$ is defined
as
\be
\frac{\dr}{\dr y^j}(\la(f_\iota\ot a^\iota))=\la (f_\iota\ot
\frac{\dr}{\dr a^j}(a^\iota)).
\ee
This definition is consistent only if $\la$ is an isomorphism. If
otherwise, there exists a non-vanishing element $f_\iota\ot
a^\iota$ such that $\la(f_\iota\ot a^\iota)=0$, but $\la
(f_\iota\ot \dr_j(a^\iota))\neq 0$.

With the condition (\ref{+20}), one classifies superfunctions as
follows \cite{bart,rog}.

(i) If the condition (\ref{+20}) is satisfied, superfunctions
(\ref{+13}) are called $GH^\infty$-superfunctions.

(ii) If $N'=0$, the condition (\ref{+20}) holds, and we deal with
$H^\infty$-superfunctions
\mar{+41}\beq
F(x,y)=\op\sum_{r=0}^m \frac1{r!}\left[
 \op\sum_{p=0}^N\frac{1}{p!}\frac{\dr^qf_{j_1\ldots j_r}}{\dr
z^{i_1}\cdots \dr z^{i_p}}(\si(x))s(x^{i_1})\cdots
s(x^{i_p})\right]y^{j_1}\cdots y^{j_r}, \label{+41}
\eeq
where $f_{j_1\ldots j_r}$ are smooth real functions
\cite{batch2,dew}.

(iii) If $N'=N$, the inequality (\ref{+20}) is not satisfied,
unless $m=0$. This is the case of $G^\infty$-superfunctions
\cite{rog80}.

Superfunctions of these three types are called smooth
superfunctions. They however are effected by serious
inconsistencies. Firstly, the odd derivatives of
$G^\infty$-superfunctions are ill defined. Secondly, a space of
values of $GH^\infty$-superfunctions changes from point to point
because a Grassmann algebra $\La$ fails to be a free
$\La'$-module. Though the $H^\infty$-superfunctions are free of
these defects, they are rather particular, namely, they are even
on $B^{n,0}$ and real on $\Bbb R^n\subset B^{n,0}$. The notion of
$G$-superfunctions overcomes these difficulties.

Let $\sG\cH_{N'}$ denote the sheaf of $GH^\infty$-superfunctions
on a supervector space $B^{n,m}$. Let us consider the sheaf of
graded commutative $\La$-algebras
\be
\sG_{N'}=\sG\cH_{N'}\op\ot_{\La'} \La.
\ee
There is its evaluation morphism
\be
\dl:\sG_{N'}\ni F\ot a\mapsto Fa\in C^{0\La}(B^{n,m})
\ee
to the ring $C^{0\La}(B^{n,m})$ of continuous $\La$-valued
functions on $B^{n,m}$. It is an epimorphism onto the sheaf
$\sG^\infty$ of $G^\infty$-superfunctions on $B^{n,m}$. A key
point is that, for any two integers $N'$ and $N''$ satisfying the
condition (\ref{+20}), there is the canonical isomorphism of
sheaves of graded commutative $\La$-algebras $\sG_{N'}$ and
$\sG_{N''}$. Therefore, given the sheaf $\cH^\infty$ of
$H^\infty$-superfunctions $F$ (\ref{+41}) on a supervector space
$B^{n,m}$, it suffices to consider the canonical sheaf
\mar{g23}\beq
\sG_{n,m}=\cH^\infty\ot \La, \qquad \dl: \sG_{n,m}\to
C^{0\La}_{B^{n,m}}, \label{g23}
\eeq
of graded commutative $\La$-algebras on $B^{n,m}$. Its sections
are called $G$-superfunctions.

It is important from the geometric viewpoint that the sheaf
$\gd\sG_{n,m}$ of graded derivations of the sheaf $\sG_{n,m}$
(\ref{g23}) is a locally free sheaf of $\sG_{n,m}$-modules of rank
$(n,m)$. On any open set $U\subset B^{n,m}$, the
$\sG_{n,m}(U)$-module $\gd\sG_{n,m}(U)$ is generated by the
derivations $\dr/\dr x^i$, $\dr/\dr y^j$ which act on
$\sG_{n,m}(U)$ by the rule.
\mar{+83}\beq
\frac{\dr}{\dr x^i}(F\ot a)=\frac{\dr F}{\dr x^i}\ot a, \qquad
\frac{\dr}{\dr y^j}(F\ot a)=\frac{\dr F}{\dr y^j}\ot a.
\label{+83}
\eeq

\begin{defi} \mar{g31} \label{g31}
Given two open subsets $U$ and $V$ of a supervector space
$B^{n,m}$, a map $\f: U\to V$ is called supersmooth if it is a set
of $n+m$ smooth superfunctions. A Hausdorff paracompact
topological space $M$ is said to be an $(n,m)$-dimensional smooth
supermanifold if it admits an atlas
\be
\Psi=\{U_\zeta,\f_\zeta\}, \qquad \f_\zeta: U_\zeta\to B^{n,m}
\ee
such that the transition functions $\f_\zeta\circ\f_\xi^{-1}$ are
supersmooth. If transition functions are $GH^\infty$- $H^\infty$-,
or $G^\infty$-superfunctions, one deals with $GH^\infty$-
$H^\infty$-, or $G^\infty$-supermanifolds, respectively.
\end{defi}

By virtue of Definition \ref{g31}, any smooth supermanifold of
dimension $(n,m)$ also carries a structure of a smooth real
manifold of dimension $2^{N-1}(n+m)$, whose atlas however
possesses rather particular transition functions. Therefore, it
may happen that non-isomorphic smooth supermanifolds are
diffeomorphic as smooth manifolds.

Similarly to the case of smooth manifolds, Definition \ref{g31} of
smooth supermanifolds is equivalent to the following one
\cite{bart,book00}.

\begin{defi} \mar{g32} \label{g32}
A smooth supermanifold is a graded local-ringed space $(M,S)$ with
an underlying topological space $M$ and the structure sheaf $S$
which is locally isomorphic to the graded local-ringed space
$(B^{n,m},\cS)$, where $\cS$ is one of the sheaves of smooth
superfunctions on $B^{n,m}$.
\end{defi}

By a morphism of smooth supermanifolds is meant their morphism as
local-ringed spaces
\be
(\vf,\Phi): (M,S)\to (M',S'), \quad \vf: M\to M', \quad
\Phi(S')=(\vf_*\circ\vf^*)(S')\subset \vf_*(S),
\ee
where $\vf_*(S)$ is the direct image of a sheaf and $\vf^*S'$ is
the inverse (pull-back) one. The condition $\vf^*(S')\subset S$
implies that $\vf:M\to M'$ is a smooth map.

The notion of a $G$-supermanifold follows Definition \ref{g32} of
smooth supermanifolds.

\begin{defi} \mar{g33} \label{g33}
A $G$-supermanifold is a graded local-ringed space $\wh
M=(M,\sG_M)$ satisfying the following conditions:

(i) $M$ is a Hausdorff paracompact topological space;

(ii) $(M,\sG_M)$ is locally isomorphic to the graded local-ringed
space $(B^{n,m},\sG_{n,m})$, where $\sG_{n,m}$ is the sheaf of
$G$-superfunctions on $B^{n,m}$;

(iii) there exists an evaluation morphism $\dl:\sG_M\to
C^{0\La}_M$ to the sheaf $C^{0\La}_M$ of continuous $\La$-valued
functions on $M$ which is locally compatible to the evaluation
morphism (\ref{g23}).
\end{defi}

In particular, the triple $\wh B^{n,m}=(B^{n,m},\sG_{n,m},\dl)$,
where $\dl$ is the evaluation morphism (\ref{g23}), is called the
standard $G$-supermanifold.

Any $GH^\infty$-supermanifold $(M,GH^\infty_M)$ with the structure
sheaf $\sG\cH^\infty_M$ is naturally extended to the
$G$-supermanifold $(M,\sG_M=\sG\cH^\infty_M\ot \La)$. Every
$G$-supermanifold defines the underlying $G^\infty$-supermanifold
$(M,\sG^\infty_M=\dl(\sG_M))$. Therefore, the underlying space $M$
of a $G$-supermanifold $(M,\sG_M)$ is provided with the structure
of a real smooth manifold of dimension $2^{N-1}(n+m)$. However, it
may happen that non-isomorphic $G$-supermanifolds have
diffeomorphic underlying smooth manifolds.

Given a $G$-supermanifold $(M,\sG_M)$, the ring $\sG_M(M)$ of
$G$-superfunctions on $M$ becomes a Fr\'echet algebra with respect
to the topology of uniform convergence of derivatives of any order
defined by the family of seminorms
\be
p_{l,K}(F)=\op\max_{q\in K, (\al_1+\cdots +\al_k)\leq l}
\|\dl((\frac{\dr}{\dr x^{i_1}})^{\al_1}\cdots (\frac{\dr}{\dr
x^{i_k}})^{\al_k}\op\sum_{r=0}^m\op\sum_{(j_1\ldots
j_r)}\frac{\dr}{\dr y^{j_1}}\cdots \frac{\dr}{\dr
y^{j_r}}F)(q)\|_\La,
\ee
where $l\in \Bbb N$ and $K$ runs over compact subsets of $M$.
Given  the standard $G$-supermanifold $(B^{n,m},\sG_{n,m})$ and an
open $U\subset B^{n,m}$, there is an isomertical isomorphism of
Fr\'echet algebras
\be
\sG_{n,m}(U)\equiv C^{\infty\La}(\si^{n,m}(U))\ot\w\Bbb R^m=
C^\infty(\si^{n,m}(U))\ot\La\ot\w\Bbb R^m,
\ee
where $C^\infty(\Bbb R^m)\ot\La\ot\w\Bbb R^m$ is provided with the
corresponding topology of uniform convergence of derivatives of
any order. As a consequence, the evaluation morphism $\dl$
(\ref{g23}) takes its values into the sheaf
$C^{\infty\La}_{B^{n,m}}$ of smooth $\La$-valued functions on
$B^{n,m}$. Accordingly, the evaluation morphism of a
$G$-supermanifold $(M,\sG_M)$ is $\dl:\sG_M\to C^{\infty\La}_M$.

By a morphism of $G$-supermanifolds is meant their morphism as
local-ringed spaces
\mar{g35}\beq
\wh\vf= (\vf,\Phi): (M,\sG_M)\to (M',\sG_{M'}) \label{g35}
\eeq
over the pull-back morphism $(\vf, \vf_*\circ\vf^*)$ of the
underlying $G^\infty$-supermanifolds. It follows that $\vf:M\to
M'$ is a smooth map of underlying smooth manifolds. Note that a
map $\vf$ is not sufficient on its own in order to determine an
even sheaf morphism $\Phi$, and additional specification $\Phi$ is
needed.

A $G$-morphism (\ref{g35}) is said to be a monomorphism (resp.
epimorphism) if $\vf$ is injective (resp. surjective) and $\Phi$
is an epimorphism (resp. monomorphism). In particular, a
$G$-monomorphism $\wh\vf:\wh M'\to\wh M$ is called a
supersubmanifold if a morphism of underlying manifolds $\vf:M'\to
M$ is a submanifold, i.e., an injective immersion. If $\vf$ is
imbedding, a supersubmanifold is called imbedded.

\section{Superbundles}

As was mentioned above, we consider superbundles in the category
of $G$-supermanifolds \cite{bart,book00}. Therefore, we start with
the definition of a product of two $G$-supermanifolds.

Let $(B^{n,m},\sG_{n,m})$ and $(B^{r,s},\sG_{r,s})$ be two
standard $G$-supermanifolds. Given open sets $U\subset B^{n,m}$
and $V\subset B^{r,s}$, we consider the presheaf
\mar{+67}\beq
U\times V\to \sG_{n,m}(U)\wh\ot \sG_{r,s}(V) \label{+67}
\eeq
where $\wh\ot$ denotes the tensor product of modules completed in
the Grothendieck topology. This presheaf yields the structure
sheaf $\sG_{n+r,m+s}$ of the standard $G$-supermanifold $\wh
B^{n+r,m+s}$. This construction is generalized to arbitrary
$G$-supermanifolds as follows.

Let  $(M,\sG_M)$ and $(M',\sG_{M'})$ be two $G$-supermanifolds of
dimensions $(n,m)$ and $(r,s)$, respectively. Their product
$(M,\sG_M) \times(M',\sG_{M'})$ is defined as the graded
local-ringed space $(M\times M', \sG_M\wh\ot \sG_{M'})$, where
$\sG_M\wh\ot \sG_{M'}$ is the sheaf constructed from the presheaf
\be
U\times U'\to \sG_M(U)\wh\ot \sG_{M'}(U')
\ee
for any open subsets $U\subset M$ and $U'\subset M'$. This product
is a $G$-supermanifold of dimension $(n+r,m+s)$ provided with the
evaluation morphism
\be
\dl: \sG_M\wh\ot \sG_{M'}\to C^{\infty\La}_{M\times M'}.
\ee
There are the canonical $G$-epimorphisms
\be
\wh\pr_1:(M,\sG_M) \times(M',\sG_{M'})\to (M,\sG_M), \qquad
\wh\pr_2:(M,\sG_M) \times(M',\sG_{M'})\to (M',\sG_{M'}),
\ee
which one can think of as being trivial superbundles.

\begin{defi} \mar{g36} \label{g36} A (locally trivial) superbundle
over a $G$-supermanifold $\wh M=(M,\sG_M)$ with a typical fiber
$\wh F=(F,\sG_F)$ is a pair $(\wh Y, \wh\pi)$ of a
$G$-supermanifold $\wh Y=(Y,\sG_Y)$ and a $G$-epimorphism
\mar{g37}\beq
\wh\pi: (Y,\sG_Y)\to (M,\sG_M) \label{g37}
\eeq
such that $M$ admits an open cover $\{U_\al\}$ together with a
family of local $G$-isomorphisms
\mar{g38}\beq
\wh\psi_\al: (\pi^{-1}(U_\al),\sG_Y|_{\pi^{-1}(U_\al)})\to
(M,\sG_M|_{U_\al})\times (F,\sG_F). \label{g38}
\eeq
\end{defi}

For any $q\in M$, by $\wh Y_q=\wh\pi^{-1}(q)$ is denoted the
$G$-supermanifold
\mar{g39}\beq
(\pi^{-1}(q), \sG_q=(\sG_Y/{\cal M}_q)|_{\pi^{-1}(q)}),
\label{g39}
\eeq
where ${\cal M}_q$ is a subsheaf of $\sG_Y$ whose sections vanish
on $\pi^{-1}(q)$. This supermanifold is a fiber of $\wh Y$ over
$q\in M$. By a section of the superbundle (\ref{g37}) over an open
set $U\subset M$ is meant a $G$-monomorphism $\wh s: (U,
\sG_M|_U)\to \wh Y$ such that $\wh\pi\circ\wh s$ is the identity
morphism of $(U, \sG_M|_U)$. Given another supermanifold $\wh Y'$
over $\wh M$, a superbundle morphism $\wh\vf:\wh Y\to \wh Y'$ is a
$G$-morphism such that $\wh\pi'\circ\wh\vf=\wh\pi$.

It is readily observed that, by virtue of Definition \ref{g36},
the underlying space $Y$ of a superbundle $\wh Y$ (\ref{g37}) is a
smooth fiber bundle over $M$ with the typical fiber $F$. A section
$\wh s$ of this superbundle defines a section $s$ of the fiber
bundle $Y\to M$, and a superbundle morphism $\wh Y\to\wh Y'$ is
given over a smooth bundle morphism $Y\to Y'$ of their underlying
spaces.

A superbundle over a $G$-supermanifold $\wh M=(M,\sG_M)$ whose
typical fiber is a superspace $(B^{n|m}, \sG_{n|m})$  (seen as the
standard $(n+m,n+m)$-dimensional $G$-supermanifold) is called a
supervector bundle. Transition functions
$\wh\rho_{\al\bt}=\wh\psi_\al\circ\wh\psi_\bt^{-1}$ of a
supervector bundle yield sheaf isomorphisms
\be
\Upsilon_{\al\bt}: \sG_{n|m}|_{U_\al\cap U_\bt}\to
\sG_{n|m}|_{U_\al\cap U_\bt}
\ee
which are described by matrices whose entries are sections of
$\sG_{n|m}|_{U_\al\cap U_\bt}$. Their evaluation
$\dl(\Upsilon_{\al\bt})$ are $GL(n|m;\La)$-valued
$G^\infty$-functions on $U_\al\cap U_\bt$. Sections of a
supervector bundle constitute a sheaf of locally free graded
$\sG_M$-modules. Conversely, let $S$ be a sheaf of locally free
graded $\sG_M$-modules of rank $(r,s)$ on a $G$-manifold $\wh M$,
there exists a supervector bundle over $\wh M$ such that $S$ is
isomorphic to the sheaf of its sections.

For instance, the locally free graded sheaf $\gd \sG_M$ of graded
derivations of $\sG_M$ defines a supervector bundle, called the
tangent superbundle $T\wh M$ of a $G$-supermanifold $\wh M$. If
$(q^1,\ldots,q^{m+n})$ and $(q'^1,\ldots,q'^{m+n})$ are two
coordinate charts on $M$, the Jacobian matrix
\mar{g41}\beq
\Upsilon^i_j=\frac{\dr q'^i}{\dr q^j} \label{g41}
\eeq
(see the prescription (\ref{+83})) provides the transition
function for $T\wh M$. It should be emphasized that the evaluation
$\dl(\Upsilon^i_j)$ of the Jacobian matrix (\ref{g41}) cannot be
written as the Jacobian matrix since odd derivatives of
$G^\infty$-superfunctions are ill-defined.

Turn now to the notion of a principal superbundle with a structure
$G$-Lie supergroup.

Let $\wh e=(e,\La)$ denote a single point with the trivial
$(0,0)$-dimensional $G$-supermanifold structure. For any
$G$-supermanifold, there are natural identifications $\wh
e\times\wh M=\wh M\times\wh e=\wh M$ we refer to in the sequel. A
$G$-supermanifold $\wh G=(G,\sG_G)$ is said to be a $G$-Lie
supergroup if there exist the following $G$-supermanifold
morphisms: a multiplication $\wh m:\wh G\times \wh G\to\wh G$, a
unit $\wh\ve: \wh e\to \wh G$, an inverse $\wh  k:\wh G\to\wh G$,
which satisfy

the associativity $\wh m\circ(\id \times \wh m)=\wh m\circ(\wh
m\times\id):\wh G\times\wh G\times\wh G\to \wh G\times\wh G\to \wh
G$,

the unit property $(\wh m\circ (\wh\ve\times\id))(\wh e\times \wh
G)= (\wh m\circ (\id\times\wh\ve))(\wh G\times \wh e)=\id \wh G$,

the inverse property $\wh m\circ (\wh k\times\id)(\wh G\times\wh
G)=\wh m\circ (\id\times\wh k)(\wh G\times\wh G )=\wh\ve(\wh e)$.

\noindent Given a point $g\in G$, let us denote by $\wh g:\wh
e\to\wh G$ the $G$-supermanifold morphism whose image in $G$ is
$g$. Then one can define the left and right translations of $\wh
G$ as the $G$-supermanifold isomorphisms
\mar{g55}\beq
\wh L_g: \wh G=\wh e\times \wh G\ar^{\wh g\times\id} \wh
G\times\wh G\ar^{\wh m} \wh G,\qquad \wh R_g: \wh G=\wh G\times
\wh e\ar^{\id\times\wh g} \wh G\times\wh G\ar^{\wh m} \wh G.
\label{g55}
\eeq

Apparently, the underlying smooth manifold $G$ of a $G$-Lie
supergroup $\wh G$ is provided with the structure of a real Lie
group, called the underlying Lie group. In particular, the
transformations of $G$ corresponding to the left and right
translations (\ref{g55}) are ordinary left and right
multiplications of $G$ by $g$.

For example, the general linear graded group $GL(n|m;\La)$ is
endowed with the natural structure of an $H^\infty$-supermanifold
of dimension $(n^2+m^2, 2nm)$ so that the matrix multiplication is
an $H^\infty$-morphism. Thus, $GL(n|m;\La)$ is a $H^\infty$-Lie
supergroup. It is trivially extended to the $G$-Lie supergroup
$\wh{GL}(n|m;\La)$, called the general linear supergroup.

A homomorphism $\wh\vf:\wh G\to\wh G'$ of $G$-Lie groups is
defined as a $G$-supermanifold morphism which obeys the conditions
\be
\wh\vf\circ\wh m=\wh m'\circ(\wh\vf\times \wh\vf). \qquad
\wh\vf\circ\wh\ve=\ve', \qquad \wh\vf\circ\wh k=\wh k'\circ\wh\vf.
\ee
In particular, an imbedded monomorphism $\wh H\to\wh G$ of $G$-Lie
supergroups is called a $G$-Lie supersubgroup of $\wh G$.
Accordingly, the monomorphism of underlying Lie groups $H\to G$ is
a Lie subgroup of $G$.

By a right action of a $G$-Lie supergroup $\wh G$ on a
$G$-supermanifold $\wh P$ is meant a $G$-epimorphism $\wh\rho:\wh
P\times \wh G\to \wh P$ such that
\be
\wh\rho\circ (\wh\rho\times\id)=\wh\rho\circ(\id\times\wh m):\wh
P\times \wh G\times \wh G\to \wh P,\qquad \wh\rho\circ (\id\times
\wh \ve)(\wh P\times \wh e)=\id\wh P.
\ee
Apparently, this action implies a right action $\rho$ of the
underlying Lie group $G$ of $\wh G$ on the underlying manifold $P$
of $\wh P$. Similarly, a left action of a $G$-Lie supergroup on a
$G$-supermanifold is defined.

For instance, a $G$-Lie group acts on itself by right and left
translations (\ref{g55}). The general linear supergroup
$\wh{GL}(n|m;\La)$ acts linearly on the standard supermanifold
$\wh B^{n|m}$ on the left by the matrix multiplication which is a
$G$-morphism.

A quotient of the action of a $G$-Lie supergroup $\wh G$ on a
$G$-supermanifold $\wh P$ is a pair $(\wh M, \wh \pi)$ of a
$G$-supermanifold $\wh M$ and a $G$-supermanifold morphism $\wh
\pi:\wh P\to\wh M$ such that: (i) there is the equality
\mar{+222}\beq
\wh\pi\circ\wh\rho=\wh\pi\circ\wh\pr_1:\wh P\times\wh G\to\wh
M,\label{+222}
\eeq
(ii) for any morphism $\wh\vf:\wh P\to \wh M'$ such that
$\wh\vf\circ\wh\rho=\wh\vf\circ\wh\pr_1$, there is a unique
$G$-morphism $\wh \g:\wh M\to\wh M'$ with $\wh\vf=\wh \g\circ
\wh\pi$. The quotient $(\wh M,\wh\pi)$, denoted by $\wh P/\wh G$,
need not exists. If it exists, its underlying space is $M=P/G$,
and there is a monomorphism of the structure sheaf $\sG_M$ of $\wh
M$ to the direct image $\pi_*\sG_P$. Since the $G$-Lie group $\wh
G$ acts trivially on $\wh M$, the range of this monomorphism is a
subsheaf of $\pi_*\sG_P$, invariant under the action of $\wh G$.
Moreover, there is an isomorphism $\sG_M\cong (\pi_*\sG_P)^{\wh
G}$ of $\sG_M$ to the subsheaf of $\wh G$-invariant sections of
$\sG_P$. The latter is generated by sections of $\sG_P$ on
$\pi^{-1}(U)$, $U\subset M$, which have the same image under the
morphisms
\be
\wh\rho^*:\sG_P|_{\pi^{-1}(U)}\to (\sG_G\wh\ot
\sG_P)|_{(\wh\pi\circ\wh\rho)^{-1}(U)},\qquad
\wh\pr_1^*:\sG_P|_{\pi^{-1}(U)}\to (\sG_G\wh\ot
\sG_P)|_{(\wh\pi\circ\wh\pr_1)^{-1}(U)}.
\ee

\begin{defi} \label{+225} \mar{+225}
A principal superbundle with a structure $G$-Lie supergroup $\wh
G$ is defined as a quotient $\wh\pi: \wh P\to \wh P/\wh G=\wh M$,
which is a locally trivial superbundle with the typical fiber $\wh
G$ such that there exists an open cover $\{U_\al\}$ of $M$
together with $\wh G$-equivariant trivialization morphisms
\be
\wh\psi_\al: (\pi^{-1}(U_\al),\sG_P|_{\pi^{-1}(U_\al)})\to
(U_\al,\sG_M|_{U_\al})\times \wh G,
\ee
where $\wh G$ acts on itself by left translations.
\end{defi}

Note that, in fact, we need only assumption (i) of the definition
of a quotient and the condition of local triviality of $\wh P$.
Apparently, the underlying smooth bundle $P\to M$ of a principal
superbundle $\wh P\to\wh P/\wh G$ is a principal smooth bundle
with the structure group $G$.

Given a principal superbundle $\wh\pi:\wh P\to\wh M$ with a
structure $G$-Lie supergroup $\wh G$, let $\wh V$ be a
$G$-supermanifold provided with a left action $\varrho: \wh
G\times \wh V\to\wh V$ of $\wh G$. Let us consider the right
action of $\wh G$ on the product $\wh P\times \wh V$ defined by
the morphisms
\mar{g51}\beq
\wh w:\wh P\times \wh V\times \wh G\ar^{\id\times \wh\kappa} \wh
P\times \wh G\times \wh V\ar^{\id\times\wh\Delta\times\id} \wh
P\times \wh G\times \wh G\times \wh
V\ar^{\wh\rho\times\wh\varrho^{-1}} \wh P\times\wh V, \label{g51}
\eeq
where $\wh\kappa:\wh V\times \wh G\to\wh G\times\wh V$ is the
morphism exchanging the factors, $\wh\Delta:\wh G\times\wh G\to\wh
G$ is the diagonal morphism, and
$\wh\varrho^{-1}=\wh\varrho\circ(\wh k\times\id)$. The
corresponding action of the underlying Lie group $G$ on the
underlying smooth manifold $P\times V$ reads
\be
(p\times v)g=(pg\times g^{-1}v), \qquad p\in P,\quad v\in V,\quad
g\in G.
\ee
Then one can show that the quotient $(\wh P\times \wh V)/\wh G$
with respect to the action (\ref{g51}) is a superbundle over $\wh
M$ with the typical fiber $\wh V$. It is called a superbundle
associated to the principal superbundle $\wh P$ because the
underlying smooth manifold $(P\times V)/G$ is a fiber bundle over
$M$ associated to the principal bundle $P$.

\section{The proof of Theorem 2}

Let $\wh\pi:\wh P\to \wh P/\wh G$ be a principal superbundle with
a structure $G$-Lie group $\wh G$. Let $\wh i:\wh H\to \wh G$ be a
closed $G$-Lie supersubgroup of $\wh G$, i.e., $i: H\to G$ is a
closed Lie subgroup of the Lie group $G$. Since $H$ is a closed
subgroup of $G$, the latter is an $H$-principal fiber bundle $G\to
G/H$ \cite{ste}. However, $G/H$ need not possesses a
$G$-supermanifold structure. Let us assume that the action
\be
\wh\rho:\wh G\times \wh H\ar^{\id\times \wh i}\wh G\times \wh
G\ar^{\wh m}\wh G
\ee
of $\wh H$ on $\wh G$ by right multiplications defines the
quotient
\mar{g80}\beq
\wh\zeta:\wh G\to \wh G/\wh H \label{g80}
\eeq
which is a principal superbundle with the structure $G$-Lie
supergroup $\wh H$. In this case, the $G$-Lie supergroup $\wh G$
acts on the quotient supermanifold $\wh G/\wh H$ on the left by
the law
\be
\wh\varrho: \wh G\times \wh G/\wh H=\wh G\times \wh\zeta(\wh G)\to
(\wh\zeta\circ\wh m)(\wh G\times \wh G).
\ee

Given this action of $\wh G$ on $\wh G/\wh H$, we have a $\wh
P$-associated superbundle
\mar{g87}\beq
\wh\Si=(\wh P\times \wh G/\wh H)/\wh G \ar^{\wh\pi_\Si} \wh M
\label{g87}
\eeq
with the typical fiber $\wh G/\wh H$. Since
\be
\wh P/\wh H=((\wh P\times \wh G)/\wh G)/\wh H= (\wh P\times \wh
G/\wh H)/\wh G,
\ee
the superbundle $\wh\Si$ (\ref{g87}) is the quotient $(\wh P/\wh
H, \wh\pi_H)$ of $\wh P$ with respect to the right action
\be
\wh\rho\circ (\id\times \wh i):\wh P\times \wh H\ar \wh P\times
\wh G\ar \wh P
\ee
of the $G$-Lie supergroup $\wh H$. Let us show that this quotient
$\wh\pi_H:\wh P\to \wh P/\wh H$ is a principal superbundle with
the structure supergroup $\wh H$. Note that, by virtue of the
well-known theorem \cite{ste}, the underlying space $P$ of $\wh P$
is an $H$-principal bundle $\pi_H: P\to P/H$.

Let $\{V_\kappa,\wh\Psi_\kappa\}$ be an atlas of trivializations
\be
\wh\Psi_\kappa : (\zeta^{-1}(V_\kappa),
\sG_G|_{\zeta^{-1}(V_\kappa)})\to
(V_\kappa,\sG_{G/H}|_{V_\kappa})\times \wh H,
\ee
of the $\wh H$-principal bundle $\wh G\to \wh G/\wh H$, and let
$\{U_\al,\wh\psi_\al\}$ be an atlas of trivializations
\be
\wh\psi_\al: (\pi^{-1}(U_\al),\sG_P|_{\pi^{-1}(U_\al)})\to
(U_\al,\sG_M|_{U_\al})\times \wh G
\ee
of the $\wh G$-principal superbundle $\wh P\to \wh M$. Then we
have the $G$-isomorphisms
\mar{g85}\ben
&& \wh\psi_{\al\kappa}=(\id \times \wh\Psi_\kappa)\circ
\wh\psi_\al: (\psi^{-1}_\al(U_\al\times\zeta^{-1}(V_\kappa)),
\sG_P|_{\psi^{-1}_\al(U_\al\times\zeta^{-1}(V_\kappa))})\to  \label{g85}\\
&& \qquad (U_\al,\sG_M|_{U_\al})\times
(V_\kappa,\sG_{G/H}|_{V_\kappa}) \times \wh H=(U_\al\times
V_\kappa, \sG_M|_{U_\al}\wh\ot \sG_{G/H}|_{V_\kappa})\times \wh H.
\nonumber
\een
For any $U_\al$, there exists a well-defined morphism
\be
&& \wh\Psi_\al:(\pi^{-1}(U_\al), \sG_P|_{U_\al}) \to (U_\al\times
G/H, \sG_M|_{U_\al}\wh\ot \sG_{G/H})\times \wh H=\\
&& \qquad (U_\al,\sG_M|_{U_\al})\times \wh G/\wh H\times\wh H
\ee
such that
\be
\wh \Psi_\al|_{\psi^{-1}_\al(U_\al\times\zeta^{-1}(V_\kappa))}
=\wh \psi_{\al\kappa}.
\ee
Let $\{U_\al,\wh\vf_\al\}$ be an atlas of trivializations
\be
\wh\vf_\al:
(\pi_\Si^{-1}(U_\al),\sG_\Si|_{\pi^{-1}_\Si(U_\al)})\to
(U_\al,\sG_M|_{U_\al})\times \wh G/\wh H
\ee
of the $\wh P$-associated superbundle $\wh P/\wh H\to \wh M$. Then
the morphisms
\be
(\wh\vf_\al^{-1}\times \id)\circ \wh\Psi_\al: (\pi^{-1}(U_\al),
\sG_P|_{U_\al})\to (\pi_\Si^{-1}
(U_\al),\sG_\Si|_{\pi^{-1}_\Si(U_\al)})\times\wh H
\ee
make up an atlas $\{\pi^{-1}_\Si(U_\al), (\wh\vf_\al^{-1}\times
\id)\circ \wh\Psi_\al\}$ of trivializations of the  $\wh
H$-principal superbundle $\wh P\to \wh P/\wh H$. As a consequence,
we obtain the composite superbundle (\ref{g21}).

Now, let $\wh i_h : \wh P_h\to \wh P$ be an $\wh H$-principal
supersubbundle of the principal superbundle $\wh P\to\wh M$. Then
there exists a global section $\wh h$ of the superbundle $\wh
\Si\to \wh M$ such that the image of $\wh P_h$ with respect to the
morphism $\wh \pi_H\circ\wh i_h$ coincides with the range of the
section $\wh h$. Conversely, given a global section $\wh h$ of the
superbundle $\wh \Si\to \wh M$, the inverse image
$\wh\pi_H^{-1}(\wh h(\wh M))$ is an $\wh H$-principal
supersubbundle of $\wh P\to\wh M$.

\section{Supermetrics}

Let us show that, as was mentioned above, the condition of Theorem
\ref{g20} hold if $\wh H$ is the Cartan supersubgroup of a
supermatrix group $\wh G$, i.e., $\wh G$ is a $G$-Lie
supersubgroup of some general linear supergroup
$\wh{GL}(n|m;\La)$.

Recall that a Lie superalgebra $\wh{\got g}$ of an
$(n,m)$-dimensional $G$-Lie supergroup $\wh G$ is defined as a
$\La$-algebra of left-invariant supervector fields on $\wh G$,
i.e., derivations of its structure sheaf $\sG_G$. A supervector
field $u$ is called left-invariant if
\be
(\id\ot u)\circ \wh m^* =\wh m^*\circ u.
\ee
Left-invariant supervector fields on $\wh G$ make up a Lie
$\La$-superalgebra with respect to the graded Lie bracket
\be
[u,u']=u\circ u' -(-1)^{[u][u']}u'\circ u,
\ee
where the symbol $[.]$ stands for the Grassmann parity. Being a
superspace $B^{n|m}$, a Lie superalgebra is provided with a
structure of the standard $G$-supermanifold $\wh B^{n+m,n+m}$. Its
even part $\wh{\got g}_0=\wh B^{n,m}$ is a Lie $\La_0$-algebra.

Let $\wh G$ be a matrix $G$-Lie supergroup. Then there is an
exponential map
\be
\xi(J)=\exp(J)=\op\sum_k \frac{1}{k'}J^k
\ee
of some open neighbourhood of the origin of the Lie algebra
$\wh{\got g}_0$ onto an open neighbourhood $U$ of the unit of $\wh
G$. This map is an $H^\infty$-morphism, which is trivially
extended to a $G$-morphism.

Let $\wh H$ be a Cartan supersubgroup of $\wh G$, i.e., the even
part $\wh{\got h}_0$ of the Lie superalgebra $\wh{\got h}$ of $wh
H$ is a Cartan subalgebra of the Lie algebra $\wh\ccG_0$, i.e.,
\be
\wh\ccG_0=\wh{\got f}_0 +\wh{\got h}_0, \qquad [\wh{\got
f}_0,\wh{\got f}_0]\subset \wh{\got h}_0, \qquad [\wh{\got
f}_0,\wh{\got h}_0]\subset \wh{\got f}_0.
\ee
Then there exists an open neighbourhood, say again $\wh U$, of the
unit of $\wh G$ such that any element $g$ of $\wh U$ is uniquely
brought into the form
\be
g=\exp(F)\exp(I), \qquad F\in \wh{\got f}_0, \qquad I\in \wh{\got
h}_0.
\ee
Then the open set $\wh U_H=\wh m(\wh U\times \wh H)$ is
$G$-isomorphic to the direct product $\xi(\xi^{-1}(U)\cap\wh{\got
f}_0)\times\wh H$. This product provides a trivialization of an
open neighbourhood of the unit of $\wh G$. Acting on this
trivialization by left translations $\wh L_g$, $g\in\wh G$, one
obtains an atlas of a principal superbundle $\wh G\to\wh H$.

For instance, let us consider a superspace $B^{n|2m}$, coordinated
by $(x^a,y^i,\ol y^i)$, and the general linear supergroup
$\wh{GL}(n|2m;\La)$ of its automorphisms. Let $B^{n|2m}$ be
provided with the $\La$-valued bilinear form
\mar{g90}\beq
\om=\op\sum_{i=1}^n (x^i x'^i) + \op\sum_{j=1}^m (y^j\ol y'^j- \ol
y^j y'^j). \label{g90}
\eeq
The supermatrices (\ref{+200}) preserving this bilinear form make
up the orthogonal-symplectic supergroup $\wh{OS}p(n|m;\La)$
\cite{fuks}. It is a Cartan subgroup of $\wh{GL}(n|2m;\La)$. Then
one can think of the quotient
$\wh{GL}(n|2m;\La)/\wh{OS}p(n|m;\La)$ as being a supermanifold of
$\La$-valued bilinear forms on $B^{n|2m}$ which are brought into
the form (\ref{g90}) by general linear supertransformations.

Let $\wh M$ be $G$-supermanifold of dimension $(n,2m)$ and $T\wh
M$ its tangent superbundle. Let $L\wh M$ be an associated
principal superbundle. Let us assume that its structure supergroup
$\wh{GL}(n|2m;\La)$ is reduced to the supersubgroup
$\wh{OS}p(n|m;\La)$. Then by virtue of Theorem \ref{g20}, there
exists a global section $h$ of the quotient $L\wh
M/\wh{OS}p(n|m;\La)\to \wh M$ which can be regarded as a
supermetric on a supermanifold $\wh M$.

Note that, bearing in mind physical applications, one can treat
the bilinear form (\ref{g90}) as {\it sui generis} superextension
of the Euclidean metric on the body $\Bbb R^n=\si(B^{n|m})$ of the
superspace $B^{n|m}$. However, the body of a supermanifold is
ill-defined in general \cite{iva86,caten}.

\section{Appendix. Nonlinear realization}

Let $G$ be a Lie group and $H$ its Cartan subgroup. The Lie
algebra $\ccG$ of $G$ is split into the sum $\ccG={\got f} +{\got
h}$ of the Lie algebra $\got h$ of $H$ and its supplement $\got f$
obeying the commutation relations
\mar{g70}\beq
[{\got f},{\got f}]\subset \got h, \qquad [{\got f},{\got
h}]\subset \got f. \label{g70}
\eeq
There exists an open neighbourhood $U$ of the unit of $G$ such
that any element $g\in U$ is uniquely brought into the form
\be
g=\exp(F)\exp(I), \qquad F\in{\got f}, \qquad I\in\got h.
\ee
Let $U_G$ be an open neighbourhood of the unit of $G$ such that
$U_G^2\subset U$, and let $U_0$ be an open neighbourhood of the
$H$-invariant center $\si_0$ of the quotient $G/H$ which consists
of elements
\be
\si=g\si_0=\exp(F)\si_0, \qquad g\in U_G.
\ee
Then there is a local section $s(g\si_0)=\exp(F)$ of $G\to G/H$
over $U_0$. With this local section, one can define the induced
representation (\ref{g4}) of elements $g\in U_G\subset G$ on
$U_0\times V$ given by the expressions
\mar{g5,6}\ben
&& g\exp(F)=\exp(F')\exp(I'), \label{g5}\\
&& g:(\exp(F)\si_0,v)\mapsto (\exp(F')\si_0,\exp(I')v). \label{g6}
\een
In the physical literature, the representation (\ref{g6}) is
called a nonlinear realization of $G$ \cite{col,jos}. The
following is the corresponding representation of the Lie algebra
$\ccG$ of $G$.

Let $\{F_\al\}$, $\{I_a\}$ be the bases for $\got f$ and $\got h$,
respectively. Their elements obey the commutation relations
\be
[I_a,I_b]= c^d_{ab}I_d, \qquad [F_\al,F_\bt]= c^d_{\al\bt}I_d,
\qquad [F_\al,I_b]= c^\bt_{\al b}F_\bt.
\ee
Then the relation (\ref{g5}) leads to the formulas
\mar{g7,7',8}\ben
&& F_\al: F\mapsto F'=F_\al
+\op\sum_{k=1}l_{2k}[\op\ldots_{2k}[F_\al,F],F],\ldots,F]-
l_n\op\sum_{n=1}
 [\op\ldots_n[F,I'],I'],\ldots,I'], \label{g7}\\
&& \qquad I'=
\op\sum_{k=1}l_{2k-1}[\op\ldots_{2k-1}[F_\al,F],F],\ldots,F],
\label{g7'}\\
&& I_a: F\mapsto F'=
2\op\sum_{k=1}l_{2k-1}[\op\ldots_{2k-1}[I_a,F],F],\ldots,F],
\qquad I'=I_a, \label{g8}
\een
where coefficients $l_n$, $n=1,\ldots$, are obtained from the
recursion relation
\mar{g71}\beq
\frac{n}{(n+1)!}=\op\sum_{i=1}^n\frac{l_i}{(n+1-i)!}. \label{g71}
\eeq
Let $U_F$ be an open subset of the origin of the vector space
$\got f$ such that the series (\ref{g7}) -- (\ref{g8}) converge
for all $F\in U_F$, $F_\al\in\got f$ and $I_a\in\got h$. Then the
above mentioned nonlinear realization of the Lie algebra $\ccG$ in
$U_F\times V$ reads
\mar{g9}\beq
F_\al: (F,v)\mapsto (F',I'v), \qquad I_a:(F,v)\mapsto (F',I'v),
\label{g9}
\eeq
where $F'$ and $I'$ are given by the expressions (\ref{g7}) --
(\ref{g8}). In physical models, the coefficients $\si^\g$ of
$F=\si^\g F_\g$ are treated as Goldstone fields. They are usually
small, and one considers the representation (\ref{g9}) up to the
of the second order in $\si^\al$. It reads
\be
&& F_\al(\si^\g)=\dl^\g_\al +
\frac{1}{12}(c^\bt_{\al\m}c^\g_{\bt\nu} - 3 c^b_{\al\m}c^\g_{\nu
b})\si^\m\si^\nu,\qquad F_\al(v)= \frac12c^b_{\al\nu}\si^\nu
I_b (v),\\
&& I_a(\si^\g)=c^\g_{a\nu}\si^\nu, \qquad I_a(v)=I_a (v).
\ee

In a similar way, the nonlinear realizations of Lie superalgebras
are constructed. Let $\wh H$ be a $G$-Lie supersubgroup of $\wh G$
such that the even part $\wh{\got h}_0$ of its Lie superalgebra is
a Cartan subalgebra of the Lie algebra $\wh\ccG_0$. With
$F,F',F''\in\wh{\got f}_0$ and $I',I''\in\wh{\got h}_0$, we can
repeat the relations (\ref{g5}), (\ref{g7}) -- (\ref{g8}) as
follows:
\mar{g72,3,4}\ben
&& F''\exp(F)=\exp(F')\exp(I'), \nonumber\\
&& \qquad F'=F''
+\op\sum_{k=1}l_{2k}[\op\ldots_{2k}[F'',F],F],\ldots,F]-
l_n\op\sum_{n=1}
 [\op\ldots_n[F,I'],I'],\ldots,I'], \label{g72}\\
&& \qquad I'=
\op\sum_{k=1}l_{2k-1}[\op\ldots_{2k-1}[F'',F],F],\ldots,F],
\label{g73}\\
&& I''\exp(F)=\exp(F')\exp(I'), \nonumber\\
&& \qquad F'=
2\op\sum_{k=1}fl_{2k-1}[\op\ldots_{2k-1}[I'',F],F],\ldots,F],
\qquad I'=I'', \label{g74}
\een
where coefficients $l_n$, $n=1,\ldots$, are obtained from the
formula (\ref{g71}).

Let a superspace $\wh V$ carries out a linear representation of
the Lie superalgebra $\wh{\got h}$. Let $\wh U_F$ be an open
subset of the supervector space $\wh{\got f}_0$ such that the
series (\ref{g72}) -- (\ref{g74}) converge for all $F\in \wh U_F$,
$F''\in\wh{\got f}_0$ and $I''\in\wh{\got h}_0$. Then we obtain
the following nonlinear realization of the even Lie algebra
$\wh\ccG_0$ in $\wh U_F\times \wh V$:
\be
F'': (F,v)\mapsto (F',I'v), \qquad I'':(F,v)\mapsto (F',I'v),
\ee
where $F'$ and $I'$ are given by the expressions (\ref{g72}) --
(\ref{g74}).

\end{document}